\documentclass[useAMS,usenatbib]{mn2e}
\usepackage{epsfig}
\newif\ifAMStwofonts

\newcommand{\uc}{{\sc ultracam}}
\newcommand{\nmus}{GRS\,1124--684}
\newcommand{\cen}{Cen\,X--4}
\newcommand{\sloanu}{$\it u'$}
\newcommand{\sloang}{$\it g'$}
\newcommand{\sloani}{$\it i'$}
\newcommand{\rtr}{$\it r_{\bf tr}$}
\newcommand{\rd}{$\it R_{\bf D}$}

\newcommand{\rsch}{$\rm R_{\bf sch}$}

\newcommand{\Rsun}{$\rm R_{\odot}$}

\title[Observations of the quiescent X-ray transients GRS\,1124--684 
(=GU\,Mus) and Cen\,X--4 (=V822\,Cen)]
{Observations of the quiescent X-ray transients GRS\,1124--684 
(=GU\,Mus) and Cen\,X--4 (=V822\,Cen) taken with \uc\ on the VLT}
\author[T. Shahbaz et al.]
       {T.\,Shahbaz,$^{1}$\thanks{E-mail: tsh@ll.iac.es}
        V.S.\,Dhillon$^2$,
        T.R.\,Marsh$^3$,
        J.\,Casares$^1$,
        C.\,Zurita$^1$,
        P.A.\,Charles$^4$\\
$^1$Instituto de Astrof\'\i{}sica de Canarias, 38200 La Laguna,
    Tenerife, Spain \\
$^2$Department of Physics and Astronomy, University of Sheffield,
    Sheffield, S3 7RH, UK  \\
$^3$Department of Physics, University of Warwick, Coventry CV4 7AL, England \\
$^4$South African Astronomical Observatory, P.O. Box 9, Observatory, 7935,
    South Africa\\
}

\begin{document}

\pagerange{\pageref{firstpage}--\pageref{lastpage}} \pubyear{2009}

\maketitle

\label{firstpage}

\begin{abstract}

\noindent
We present high time-resolution multicolour optical observations of the 
quiescent X-ray transients \nmus\ (=GU\,Mus) and \cen\ (=V822\,Cen)  obtained
with \uc.  Superimposed  on the secondary stars' ellipsoidal modulation in both
objects  are large flares on time-scales  of 30--60\,min, as well as several
distinct rapid flares on time-scales of  a few minutes, most of which show 
further variability and unresolved structure.  Not significant quasi-periodic
oscillations are observed and the power density spectra of \nmus\ and \cen\ can
be described by a power-law.  From the colour-colour diagrams of the flare
events, for \nmus\  we find that the flares can be described by hydrogen gas
with a density of $N_{\rm H}\sim10^{24}$\,nucleons cm$^{-2}$, a  temperature of
$\sim$8000\,K and arising from a radius of $\sim$0.3\,\Rsun. Finally we compile
the values for the transition radius (the radius of the hot advection-dominated
accretion flow) estimated from quasi-periodic oscillations and/or breaks in the
power density spectrum for a variety of X-ray transients in different X-ray
states. As expected, we find a strong correlation between the bolometric
luminosity and the transition radius.

\end{abstract}


\begin{table*}
\caption{Log of the VLT+ULTRACAM observations. }
\begin{center}
\begin{tabular}{lccccc}\hline 
Object  & UT Date  & UT Start & exp. time (s) & No.of images & Median seeing (range) \\
        & starting         &         & \sloanu/\sloang/\sloani &  
	\sloanu/\sloang/\sloani & (\arcsec) \\
\hline
 \nmus   &  09/05/05 & 23:13 & 5.0/5.0/5.0  & 4005 & 1.5 (1.2--2.0) \\
 \nmus   &  10/05/05 & 00:17 & 5.0/5.0/5.0  & 3469 & 1.7 (1.3--3.0) \\
 \cen    &  18/06/07 & 01:41 & 6.0/3.0/3.0  & 3296/6579/6579/ & 0.6 (0.45 --1.2) \\ \hline	 
\end{tabular}
\end{center}
\label{log}
\end{table*}

\section{Introduction}
\label{sec:intro}
\alph{footnote} \protect\footnotetext[1]{Based on observations made at
the European Southern Observatory, Paranal, Chile (ESO program
075.D-0193)} 

X-ray transients (XRTs)  are  a subset of  low-mass X-ray  binaries 
that  display episodic, dramatic X-ray  and optical  outbursts,  usually
lasting for several months. More than 70 percent of XRTs are  thought to
contain black holes \citep{Charles06}. The black hole X-ray transients are
known to exhibit five distinct X-ray  spectral states, distinguished by the
presence or absence of a soft blackbody component at 1\,keV and the luminosity
and spectral slope of emission at harder energies; these are known as  the
quiescent, low, intermediate, high and very high states \citep{Tanaka96}.  {The
quiescent and low states can mostly  be explained with the advection dominated
accretion flow (ADAF) model  (Narayan, McClintock \& Yi 1996; Esin, McClintock
\& Narayan 1997).  In the context of the ADAF model,   properties similar to
the low/hard state are expected for the  quiescent state, as there is thought
to be  no distinction between the two except that the mass accretion rate is
much higher and the size of the ADAF region is smaller for the low/hard state. 

Similar to the transition between the low/hard and high/soft
(thermal-dominant)  state, where there  is a  reconfiguration of the accretion
flow  \citep{Esin97},  there is also observational evidence for a state
transition between the low/hard and quiescent  states,  in that the quiescent
state power-law appears softer than in the  low/hard state. In both these
states, the ADAF model predicts  that the inner edge of the disc is truncated
at some large radius,  with the interior region filled by an ADAF. Strong
evidence for such a  truncated disc is provided by observations of
XTE\,J1118+480 in the low/hard state  during outburst (\citealt{Hynes00};
\citealt{McClintock01}; \citealt{Esin01}; \citealt{Chaty03}), where the disc
has  an inner radius   of $>$55 Schwarzschild radii (\rsch) and a hot
optically-thin plasma in the inner regions. In quiescence, the ADAF model
predicts that the inner disc edge will move  outward to larger radii
\citep{Esin97}. 

Variability is one of the ubiquitous characteristics of accreting black holes
extending across all wavelengths. High time resolution optical photometry of
XRTs in quiescence have shown variability with a power density spectrum (PDS)
very similar to those of low/hard-state XRTs, i.e. a power-law band-limited
spectrum \citep{Hynes03G}. The similarity of the PDS suggests that the optical
variability could have a similar origin and might be associated with the
central X-ray source.  Furthermore the presence of optical/X-ray correlations
\citep{Hynes04} requires some association between the optical and the source of
the X-rays. The origin of the break in the band-limited
spectrum is not known, but it is plausible to expect
that it approximately scales with the size of the inner region. A low break
frequency in quiescence would then be expected, since the advective region is
expected to be larger.  Here we report on our high-time resolution multi-colour
optical observations of the black hole X-ray transient \nmus\ and the neutron
star X-ray transient  \cen\  in quiescence.  These observations are part of a
ongoing campaign with \uc\ to obtain high-time resolution photometry of X-ray
binaries (\citealt{Shahbaz03}; \citealt{Shahbaz05}).

\section{Observations and Data Reduction}
\label{sec:obs}

Multi-colour photometric observations of \nmus\ and \cen\ were obtained with
\uc\ on the 8.2-m MELIPAL unit of the Very Large Telescope (VLT) at Paranal,
Chile. The \nmus\ data were taken during the nights starting  2005 May 9 to 10,
whereas the  \cen\ data were taken on the nights starting 2007 June 18  (see
Table \ref{log} for a log of the observations). \uc\ is an ultra-fast,
triple-beam CCD camera, where the light is split into three  broad-band colours
(blue, green and red) by two dichroics. The detectors are back-illuminated,
thinned, E2V frame-transfer 1024$\times$1024 CCDs  with a pixel scale of
0.15\,arcsecs/pixel. Due to the architecture of the CCDs the dead-time is
essentially zero (for further details see \citealt{Dhillon07}).  Our
observations were taken using the Sloan \sloanu, \sloang\ and \sloani\ filters
with effective wavelengths of 3550\,\AA\,, 4750\,\AA\ and 7650\,\AA\
respectively. 


The \uc\ pipeline reduction procedures were used to debias and  flatfield  the
data. The same pipeline was also used to obtain lightcurves for \nmus,  \cen\ 
and several comparison stars by extracting the counts using aperture photometry.
The most  reliable results were obtained using a variable aperture which scaled
with the seeing.  The count ratio of the target with respect to a bright local
standard (which has similar colour to our target) was then determined.   The
magnitude of the target was then obtained using the calibrated magnitude of the
local standard. As a check of the photometry and systematics in the reduction
procedure, we also extracted lightcurves of a faint comparison star similar in
brightness to the target.  

The mean magnitude of \nmus\ was \sloanu=21.49, \sloang=20.65 and \sloani=19.92
and  we estimate the photometric accuracy per exposure to be 16.4, 2.8 and 2.0
percent  for the \sloanu, \sloang\ and \sloani\ band respectively. For \cen, the
mean magnitude was \sloanu=18.69, \sloang=18.07 and   \sloani=17.27 and  we
estimate the photometric  accuracy per exposure to be 5.9, 0.9 and 0.6 percent
for the \sloanu, \sloang\ and \sloani\ band respectively. The uncertainties only
reflect the 1-$\sigma$  statistical errors in the relative photometry.

\section{Short-term variability}
\label{sec:var}

The optical lightcurves of \nmus\ and \cen\  (Figure \ref{lcurve}) show
short-term variability/flares superimposed on  the secondary star's weak
ellipsoidal modulation. So if we want to determine the flux of the flares,
these ``steady'' contributions  must first be removed from the lightcurves. In
order to isolate the short-term variability in each band we first de-reddened
the observed magnitudes using a colour excess of E(B-V)=0.29 \citep{Cheng92}
for \nmus\ and E(B-V)=0.10 for \cen\ \citep{Blair84}  and adopting the ratio 
$A_{\rm V}/E(B-V)$=3.1 \citep{Cardelli89}, and then converted the Sloan AB
magnitudes to flux density \citep{Fukugita96}. We then fitted a double sinusoid
to the lower-envelope of the   lightcurve with periods equal to the orbital
period  [$P_{\rm orb}$=0.4320604\,d for \nmus; \citet{Casares97} and  $P_{\rm
orb}$=0.6290496\,d for \cen; \citet{Torres02}]  and its first harmonic, where
the phasing was fixed and the normalizations were allowed to float free. We
rejected points more than 3-$\sigma$ above the fit, then re-fitted,  repeating
the procedure until no new points were rejected.   For \nmus, contrary to
\cite{Hynes03G} we do not find evidence of any phase shift  in the data,
because one can see from Figure \ref{lcurve}  that orbital phase 0.0 coincides
with the  minimum light.

Figures \ref{flares_nmus} and \ref{flares_cen} show the resulting  lightcurves
(after subtracting the secondary star's ellipsoidal modulation), where large
flare events lasting 30-60\,min and 10--30\,min respectively are clearly
seen.  For \nmus\  what is also noticeable are numerous rapid flare events
(Figure \ref{profiles}), that typically last 5\,min or less and which are not
resolved.   For both targets there are numerous short-term flares with
rise/decay times of  $\sim$3--5\,min, superimposed on larger flares. For
\nmus\ the large flares have time-scales of 30--60\,min, whereas for \cen\ they
typically have time-scales of 10--30\,min. The parameters of the flares as
defined by \citet{Zurita02} are given in Table \ref{Table:Flares}. 

A look at the lightcurves of \nmus\ shows short-term  flare events which seem
to occur regularly. In Figure \ref{pulse} we show  tick marks with
representative intervals of 6.3\,min. As one can see, some of the flare  events
seem to recur on a regular basis and the period appears to be  stable on short
time-scales. It is clear, however, that these events are not strictly periodic,
and not strong enough to show up as QPOs in a log-log plot of the PDS (see
section \ref{sec:pds}).

\begin{table*}
\caption{Properties of the \nmus\ and  \cen\ flares. 
$v_{obs}$ is the spectroscopic
veiling, $v^{'}_{d}$ is the contribution to the  {\it non-variable}
disc light, $\bar{z}_{f}$ and $\sigma_{z}$ are the mean flare flux and
its standard deviation, respectively.  $\sigma^{*}_z = \sigma_z /
v^{'}_{d}$  and  $\eta$ is  the fraction of the average veiling due to
the flares \citep{Zurita02}.}
\begin{center}
\begin{tabular}{llcccccc} \hline Target & Band  & $v_{obs}$  &
$v^{'}_{d}$ & $\bar{z}_{f}$  & $\sigma_z$ &  $\sigma^{*}_z$ & $\eta$
\\ \hline 
\noalign{\smallskip} 

\nmus\   &\sloanu &  94\%  &   0.92  & 0.387 &  0.25   &  0.27  &  0.27 \\  
         &\sloang &  64\%  &   0.57  & 0.194 &  0.16   &  0.28  &  0.25 \\  
         &\sloani &  15\%  &   0.04  & 0.133 &  0.11   &  2.8   &  0.78 \\  
\\
\cen\    &\sloanu &  60\%  &   0.48  & 0.312 &  0.13   &  0.27  &  0.39 \\  
         &\sloang &  20\%  &   0.78  & 0.182 &  0.12   &  0.15  &  0.19 \\  
         &\sloani &   5\%  &   0.70  & 0.042 &  0.005  &  0.007 &  0.06 \\  
\noalign{\smallskip}
\hline
\end{tabular}
\end{center} 
\label{Table:Flares}
\end{table*}

\section{The power density spectrum}
\label{sec:pds}

The flare lightcurves (i.e. the after subtracting the secondary star's
ellipsoidal modulation from the de-reddened lightcurves) of \nmus\ and \cen\
show features which seem to be periodic.  To see if this is true we computed
the power density spectrum (PDS) of the data. To compute the PDS we detrended
the data using the double sinusoid fit described in the previous section and
then added the mean  flux level of the data.    Although the \uc\ sampling is
perfectly uniform over short periods of time (tens of minutes),  we use the
Lomb-Scargle method   to compute the periodograms} \citep{Press92} with the
same normalization method as is commonly used in  X-ray astronomy, where the
power is normalized to the fractional root mean amplitude squared per hertz
\citep{Klis94}.  We used the constraints imposed by the Nyquist frequency  and
the typical duration of each observation to limit the range of different
frequencies searched.

\begin{figure*}
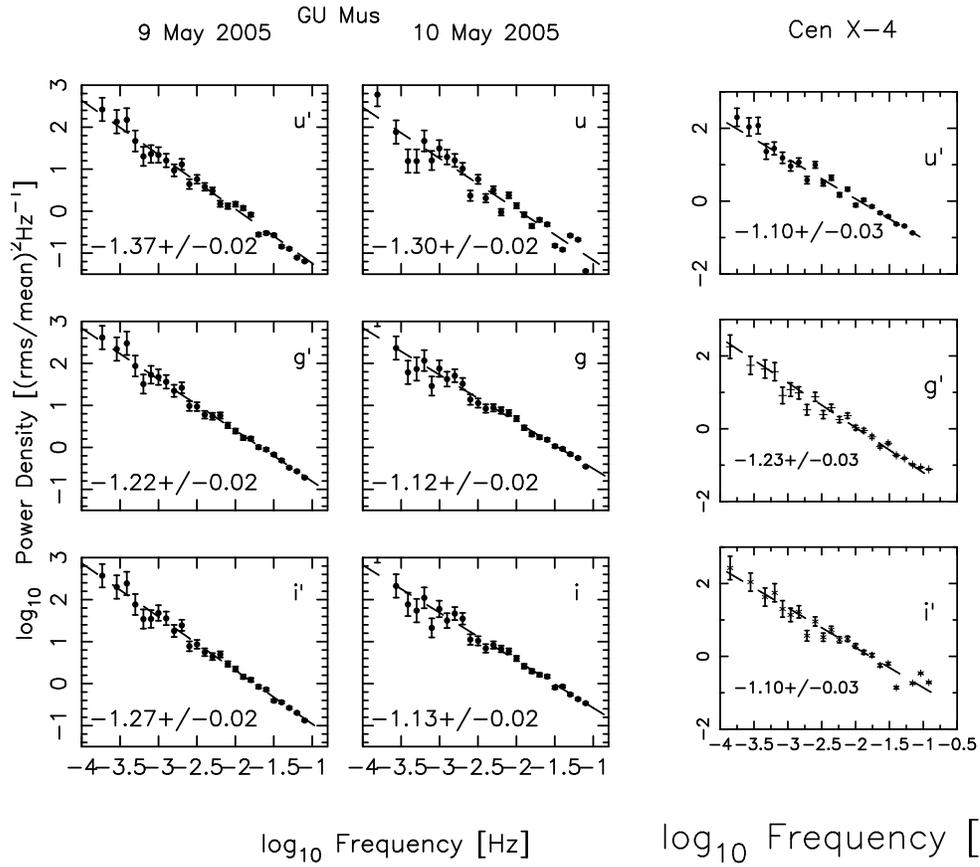

  \psfig{angle=0,width=8.0cm,file=shahbaz_fig6a.eps}
  \hspace{5mm}
  \psfig{angle=0,width=5.0cm,file=shahbaz_fig6b.eps}
  \caption{From top to bottom: the \sloanu, \sloang\ and \sloani\ PDS
of the flare lightcurves of \nmus\ taken on 9 and 10 May
2005 (left) and \cen\ taken on 18 June 2007 (right). The  dashed line is a  power--law  fit. The slope of the power-law
fit to the PDS  is indicated  in each panel.}
  \label{pds} 
\end{figure*}

The lightcurves  of \nmus\ and \cen\ shows features that are present in all
three bands (see Figure \ref{pds}).  In order to determine the significance of
possible peaks above the red-noise level, we  used a  Monte Carlo simulation
similar to \citet{Shahbaz05}. We generated lightcurves with exactly the same
sampling and integration times as the real data. We started with a model noise
lightcurve  generated  using the method of \citet{Timmer95} with a  power-law
index as determined from the PDS of the observed data, and  then added Gaussian
noise  using the errors derived from the  photometry.  We computed 5000
simulated lightcurves and then calculated the 95\% confidence level at each
frequency  taking into account a  realistic number of independent trial
\citep{Vaughan2005}.  For \nmus\ and \cen\ the 95\% confidence level contour rules out any QPOs and
indeed it  would seem that the PDS is dominated by red-noise. Thus the 
PDS of the flare lightcurve for \nmus\ and \cen\   in each band can be described by a
power-law model (see Figure \ref{pds}). 

In Figure \ref{pds} the errors in each frequency bin are  determined from the
standard deviation of the points within each bin and the  white noise level was
subtracted by fitting the highest frequencies with a  white-noise (constant)
plus red-noise (power-law) model. The power-law index of the fit is also
given. 

\section{The origin of the flares}
\label{sec:flares}

In an attempt to interpret the broad-band spectral properties of the flare,  
we compared the observed colours with the prediction for  three different 
emission mechanisms, namely a blackbody, an optically-thin layer of hydrogen
and synchrotron emission.  We computed the given emission spectrum  and then
calculate the expected flux density ratios $f_{u'}/f_{g'}$  and $f_{g'}/f_{i'}$
using the synthetic photometry  package {\sc SYNPHOT} ({\sc iraf/stsdas}). 
Given the intrinsic model flux we  can then determine the  corresponding radius
of the region that produces the observed de-reddened flux at a given distance. 

\begin{figure*}
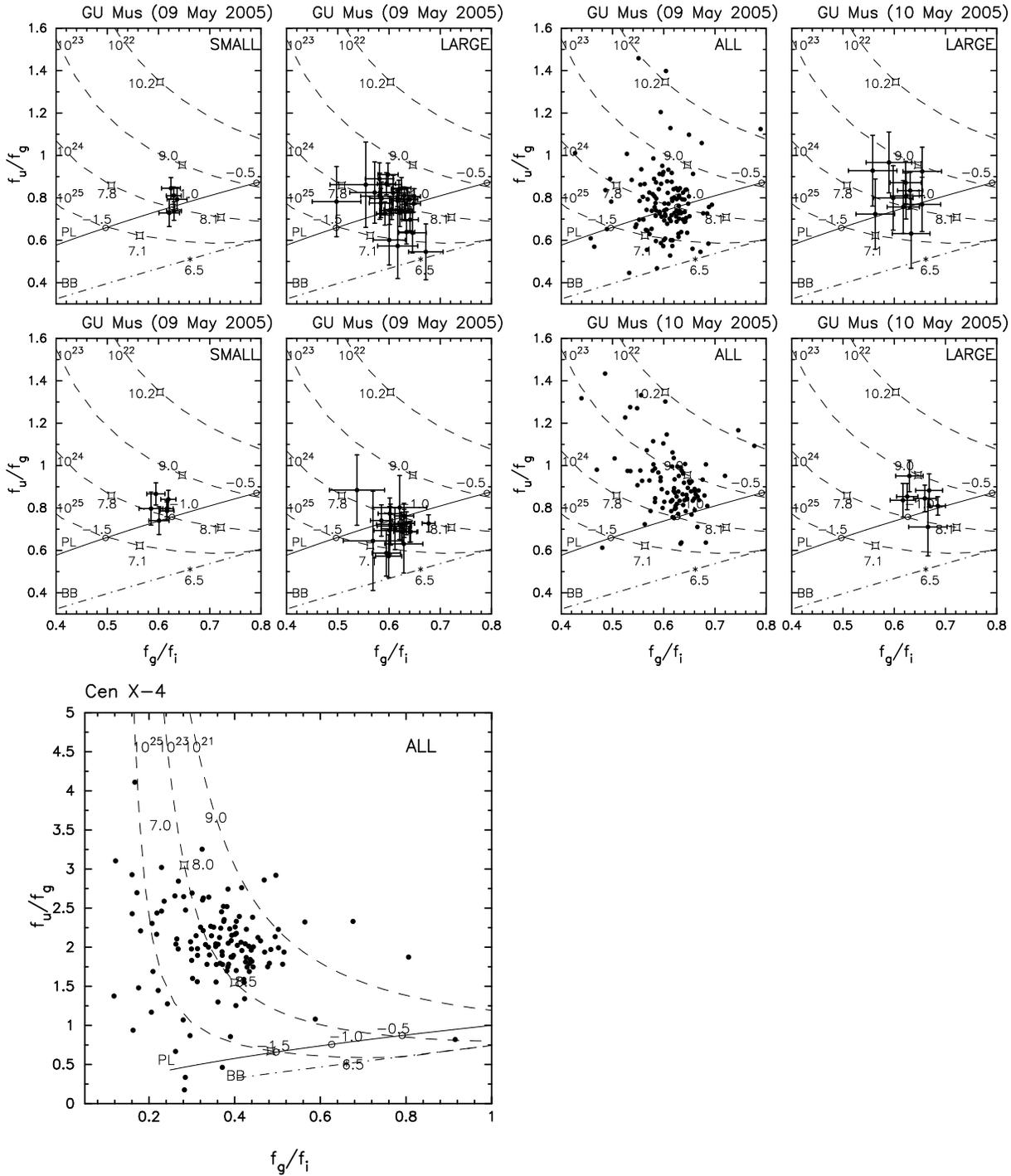

\begin{center}
  \psfig{angle=0,width=8.0cm,file=shahbaz_fig7a.eps}
  \psfig{angle=0,width=8.0cm,file=shahbaz_fig7b.eps}
\end{center}

  \hspace*{-80mm}
  \psfig{angle=0,width=7.5cm,file=shahbaz_fig7c.eps}

  \caption{Colour-Colour diagram for the clearest small and large
flare events in \nmus\ and the large flare events in \cen. The panels labelled 
"ALL" show all the data for that particular night, 
where the error bars have been
removed for clarity.
The dashed lines show optically-thin hydrogen slab models 
for  different column densities and the
open squares shows the temperature in 1000\,K units. The solid line shows
a power-law  model ($F_{\nu} \propto \nu^{\alpha}$) with indices
$\alpha$=-0.5 and -1.5 marked as open circles. The dot-dashed line is
a blackbody model where the asterisk shows the  colour for a temperature of
6500\,K.   }
  \label{colcol} 
\end{figure*}

Figure \ref{colcol} show the data for the flare events and the expected 
results for different emission models.  For stellar flares the power-law index
of the electron energy distribution $\sim$--2 \citep{Crosby93}, which
corresponds to a spectral energy distribution with index $\alpha$=--0.5 
($F_{\nu} \propto \nu^{\alpha}$). However, the spectral energy distribution
index observed in V404\,Cyg  ($\alpha \sim$ --2.0) implies a much steeper index
for the  electron energy distribution \citep{Shahbaz03}, which may not be
completely implausible given  the extreme conditions around a black hole.
Therefore in Figure \ref{colcol}  we show the power-law  model for $\alpha$
ranging from -2 to 2. We also show the very unlikely blackbody case, where the
flares are due to blackbody radiation from  a heated region of the disc's
photosphere. The most likely model for a thermal flare is emission from an 
optically thin layer of recombining hydrogen, which is essentially the 
mechanism generally accepted for solar flares. We therefore  determined the
continuum emission spectrum of an LTE slab of hydrogen for different baryon
densities,      $N_{\rm H}=10^{21}-10^{24}$\,nucleons cm$^{-2}$, and
calculated  the expected flux ratios.

For \cen\ the amplitudes of the flares are small and difficult to define. All
we can say is that they arise from an optically thin region with  $N_{\rm
H}\sim10^{21-25}$\,nucleons cm$^{-2}$, a temperature of $\sim$8,300\,K and a
radius of $\sim$0.06\,\Rsun\ (assuming a distance of 1.2\,kpc;
\citealt{Chevalier89}).   For \nmus\ we can comment more because the flares are
well defined; Figure \ref{colcol} shows the data for the best resolved small
and large  flares  for \nmus. The flare events can be described  by hydrogen 
gas with a density of
$N_{\rm H} \sim 10^{24}$\,nucleons cm$^{-2}$ and a temperature of
$\sim$8000\,K, which corresponds to a radius of $\sim$ 0.3\,\Rsun\ (assuming a
distance of 5.1\,kpc; \citealt{Gelino01}).

\section{Discussion}
\label{sec:dis}

\subsection{The origin of the flares}

We can compare the flares in \nmus\ to the flare properties determined in
other  quiescent black hole XRTs.   The large flares in V404\,Cyg are
consistent with optically thin gas with a temperature of $\sim$8000\,K arising
from a region with an equivalent blackbody radius of at least 2\,\Rsun\
\citep{Shahbaz03}. One should regard the equivalent radius estimate as a  lower
limit, because the emission mechanism is unlikely to be blackbody.  Similarly
for XTE\,J1118+480, the short-term flares have a blackbody temperature of
$\sim$3500\,K and an equivalent radius of $\sim$0.10\,\Rsun.  By isolating the
spectrum of the flare event in A0620--00, we found that it could be represented
by optically thin gas of hydrogen with radius 0.04\,\Rsun\ and temperature
10,000--14,000\,K \citep{Shahbaz05}, which  are consistent with the bright spot
area and temperature.  The Balmer line flux and variations in A0620--00
suggests that there are two emitting regions, the accretion disc and the
accretion stream/disc impact region. The persistent emission is optically thin
and during the flare events there is either an increase in temperature or the
emission is more optically thick than the continuum.  For \nmus\ we find that
the flare events reach a maximum radius of $\sim$0.3\,\Rsun, much larger than
what is observed in A0620--00 \citep{Shahbaz05}, XTE\,J1118+480
\citep{Shahbaz04} and \cen\  (section \ref{sec:flares}). 

The flares \nmus\ and \cen\  arise from various  regions in the accretion disc
that in total occupy 3.4\% and 0.3\% of the disc's area respectively. Thus
could be from the hotspot, but the  H$\alpha$ Doppler maps of \nmus\ and \cen\
do not show any evidence of a hotspot (\citealt{Casares97}; 
\citealt{Torres02}). However, it should be noted that the optical state of
quiescent transients are known to vary significantly from epoch to epoch
\citep{Cantrell08}.

\subsection{The transition radius}

\citet{Narayan97} have shown that the X-ray observations of quiescent XRTs can
be explained by a two-component accretion flow model. The geometry of the flow
consists of an hot inner ADAF that extends from the black hole horizon to a
transition radius \rtr\ and a thin accretion disc that extends from \rtr\ to
the  outer edge of the accretion disc.  An ADAF has turbulent gas at all radii,
with a variety of time-scales, ranging from a slow time-scale at the transition
radius down to nearly the free-fall time close to the compact object. In
principle, interactions between the hot inner ADAF and the cool, outer thin
disc, at or near the transition radius, can be a source of optical variability,
due to synchrotron emission by the hot electrons in the ADAF \citep{Esin97}.
For an ADAF the variability could be quasi-periodic and would have a
characteristic time-scale given by a multiple of the Keplerian rotation period
at \rtr.  The ADAF model also predicts that \rtr\ should get  larger as the
inner disc evaporates during the decline from outburst.

To see if there is any observational evidence for this, we have compiled  the
values for \rtr\ estimated from spectral energy distribution (SED) fits,  QPOs
and/or breaks in the PDS for a  variety of XRTs in different X-ray states (see
Table \ref{rt}).  For J1118+480 during outburst there are several estimates for
\rtr. The low/hard state X-ray PDS of J1118+480 (April 2000) shows a
low-frequency break $f_{\rm break}$ at 23\,mHz and a QPO at $\sim$80\,mHz 
\citep{Hynes03J}, which is approximately consistent with the relation observed
in other sources \citep{Wijnands99}, most likely related to an instability in
the accretion  flow that modulates the accretion rate.  The QPO frequency is
most likely related to \rtr, which can be a source of quasi-periodic
variability and \rtr\ can be estimated assuming the QPO frequency is the
Keplerian rotation period at \rtr.  The observed QPO frequency corresponds to
\rtr=670 ($t_K = 2\pi R/ v_K = 2\pi(GM/R^{3})^{-1/2} \sim
(M/10\,M_{\odot})(r/100)^{3/2}$\,s, where $R$ is the absolute radius and $r$ is
in units of \rsch).

\begin{table*}
\caption{The transition radius for various systems}
\label{rt}
\begin{center}
\begin{tabular}{lccccl}
\hline
\noalign{\smallskip}
Object      & State      & Method    & $\log[R_{\bf tr}/R_{\bf D}]$ 
& $\log[L_{\rm X}/L_{\rm EDD}]$      & Reference	      \\ \hline
GU\,Mus     & ultrasoft  & SED       & -2.92 & -0.16 & \cite{Misra99}	  \\
J1118+480   & low/hard   & SED       & -1.91 & -2.79 & \cite{Chaty03}     \\
J1118+480   & low/hard   & QPO       & -1.72 & -2.79 & \cite{Hynes03J}    \\
J1118+480   & low/hard   & PDS break & -0.77 & -2.79 & \cite{Hynes03J}    \\
J1655--40   & quiescence & Delay     & -1.51 & -5.25 & \cite{Hameury97}   \\
V404\,Cyg   & quiescence & QPO       & -1.29 & -4.89 & \cite{Shahbaz03}   \\
A0620--00   & quiescence & PDS break & -0.05 & -7.16 & \cite{Hynes03G}    \\
J1118+480   & quiescence & PDS break & -0.07 & -8.43 & \cite{Shahbaz05}   \\
\noalign{\smallskip}
\hline
\end{tabular}
\end{center}
\end{table*}

For a cellular-automaton ADAF disc \citep{Takeuchi97} , the break-frequency is
determined by the maximum peak intensity of the X-ray shots which is on the
order of the size of the advection-dominated region,  and corresponds to  the
inverse of the free-fall time-scale of the largest avalanches (Takeuchi,
Mineshige \& Negoro 1995)  at \rtr. However, it should be noted that the break
frequency depends not only on the size of the ADAF region but also on the
propagation speed of the perturbation (Mineshige priv. comm.). Since the
perturbation velocity should be less than the  free-fall velocity, the free-fall
velocity gives an upper limit to the size of the ADAF region.  Thus equation 13
of \citet{Takeuchi95} gives an upper limit to \rtr\ 
($r_{\bf tr}  < 10^{3.2} ( f_{\rm break} M_{\rm X})^{-2/3}$). 
Thus the break frequency observed in J1118+80 during outburst corresponds to
\rtr$<$5250. The SED fits give  \rtr$\sim$350,  which is similar to value
determined from the QPO to within a factor of 2; note that the variability/QPO
could have a characteristic time-scale given by a multiple of the Keplerian
rotation period at \rtr. 

\begin{figure*}
  \begin{center}
    \psfig{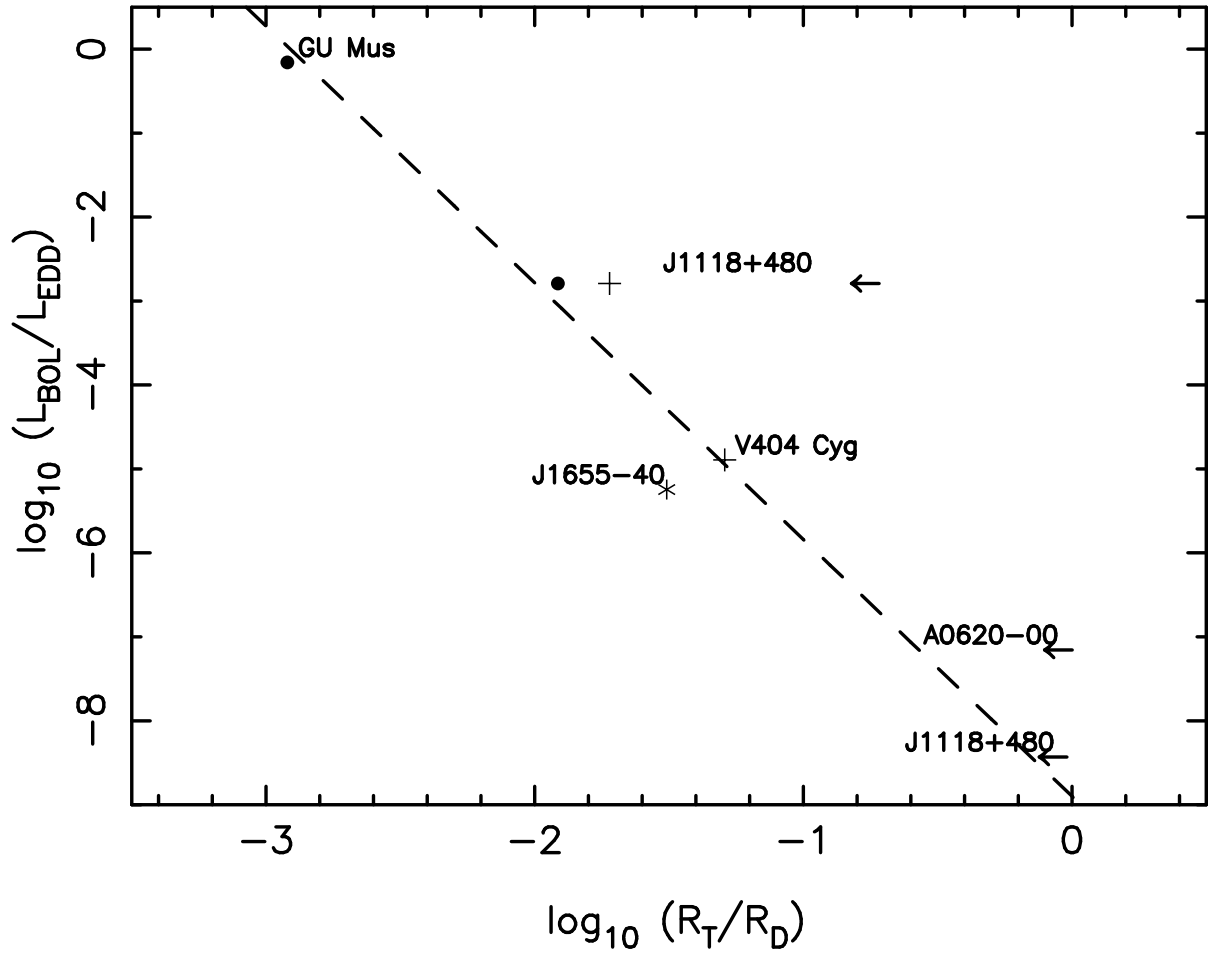}
  \end{center}
  \caption{The transition radius (accretion disc units \rd) versus 
  bolometric luminosity (Eddington units $L_{\rm EDD}$) for a variety
of X-ray transients in different X-ray states. 
The filled circles are for
systems in quiescence and the crosses are for systems in the
low/hard state or in outburst. The \rtr\ values have been determined from 
QPOs (crosses), breaks in the PDS (upper limit), fits to the spectral energy 
distribution (squares) or delays in the optical/X-ray outburst (stars): 
J1118+480 (\citealt{Chaty03};  \citealt{Shahbaz05});
A0620--00 \citep{Hynes03G}; J1655--40 \citep{Hameury97};  
V404\,Cyg \citep{Shahbaz03};  \nmus\ \citep{Misra99}. 
The X-ray luminosities are taken from \citet{Campana00} and the orbital 
parameters from \citet{Charles06}. }
  \label{lxrt} 
\end{figure*}

Figure \ref{lxrt} shows \rtr\ versus the bolometric X-ray luminosity
\citep{Campana00} for the sources listed in Table \ref{rt}. One can see that 
there does exist a correlation (the correlation coefficient  is -0.96) with a
slope index of $\sim$-3.0.  However, more observations of \rtr\ during
different X-ray states are required.

\section*{Acknowledgments}

TS, JC and CZ acknowledge support from the Spanish Ministry of Science and
Technology  under the grant AYA\,2007--66887. 
Partially funded by the Spanish MEC under the
Consolider-Ingenio 2010  Program grant  CSD2006-00070: ``First Science
with the GTC''  (http://www.iac.es/consolider-ingenio-gtc/).  
ULTRACAM is supported by STFC grant PP/D002370/1.

\newpage

\begin{figure*}
\hspace*{-80mm}
    \psfig{angle=-90,width=9.0cm,file=shahbaz_fig1a.eps}
    
\hspace*{-80mm}
    \psfig{angle=-90,width=9.0cm,file=shahbaz_fig1b.eps}
    
\hspace*{-80mm}
    \psfig{angle=-90,width=9.0cm,file=shahbaz_fig1c.eps}
 
\vspace*{-135mm}
\hspace*{100mm}
    \psfig{angle=0,width=6.5cm,file=shahbaz_fig1d.eps}

\caption{The de-reddened \sloanu, \sloang\ and \sloani-band lightcurves  of
\nmus\ and \cen\ phase-folded using the ephemeris given in \citet{Casares97} and
\citet{Torres02} respectively; orbital  phase 0.0 is  defined as inferior
conjunction of the secondary star.    The solid line is the fitted ellipsoidal
modulation. At the bottom of  each panel we also show the lightcurve of a
comparison star of similar  magnitude to the target, offset  vertically for
clarity.}
\label{lcurve} 
\end{figure*}

\begin{figure*}
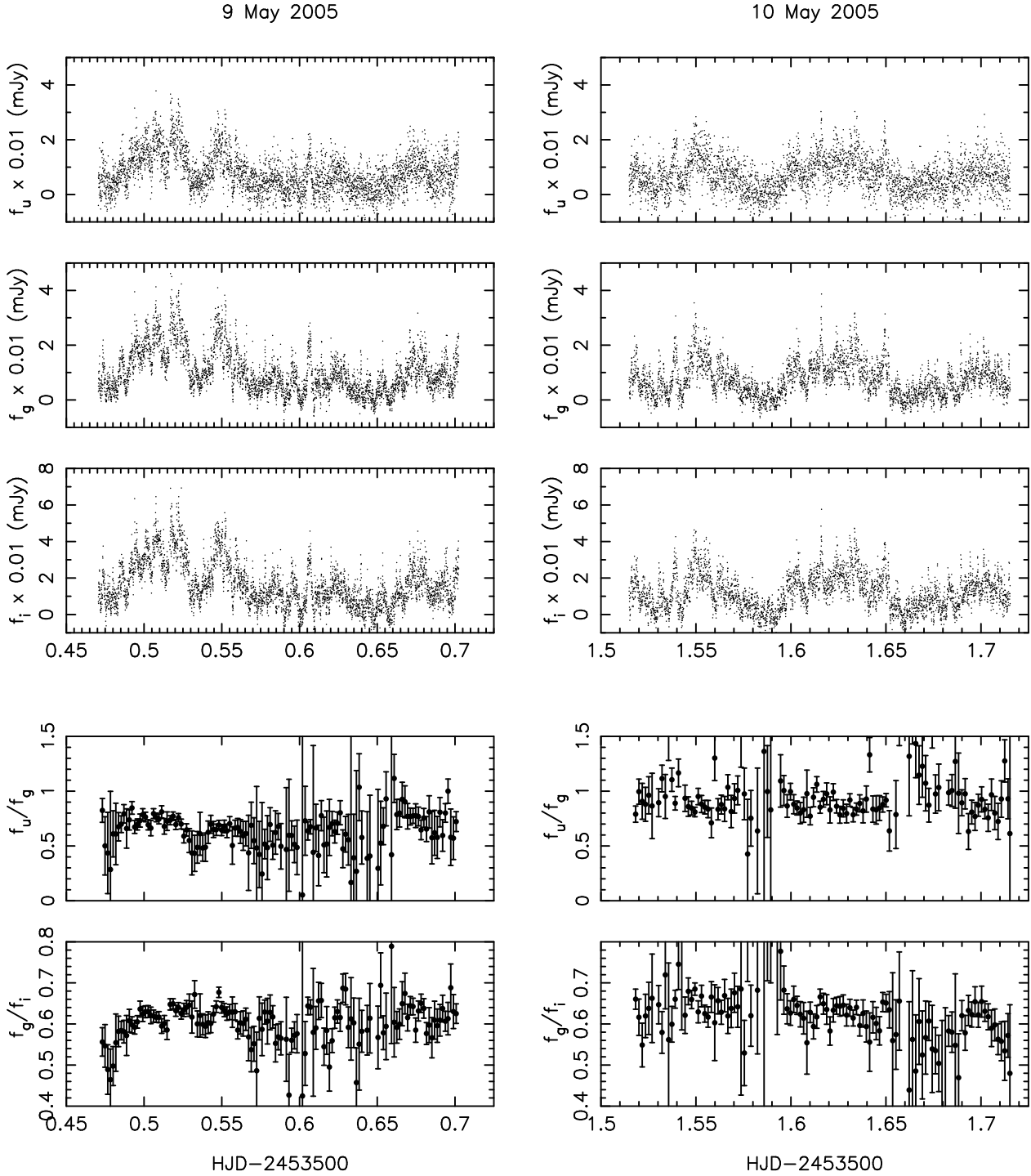

\psfig{angle=0,width=8.0cm,file=shahbaz_fig2a.eps}
\hspace{5mm} 
\psfig{angle=0,width=8.0cm,file=shahbaz_fig2b.eps}

\vspace*{10mm} 
\psfig{angle=0,width=8.0cm,file=shahbaz_fig2c.eps}
\hspace{5mm} 
\psfig{angle=0,width=8.0cm,file=shahbaz_fig2d.eps}

\caption{ Top: The \nmus\ flare de-reddened flux density \sloanu\ (top panel),
\sloang\ (middle  panel) and \sloani-band (bottom panel) lightcurves obtained by 
subtracting a fit to the lower-envelope of the lightcurves shown in
Figure \ref{lcurve}. The  uncertainties in the \sloanu, \sloang\ and \sloani\
lightcurves are   4.3$\times 10^{-3}$\,mJy, 1.2$\times 10^{-4}$\,mJy and 
2.0$\times 10^{-3}$\,mJy respectively. Bottom: the flux ratio lightcurves binned
to a time resolution of 150\,s for clarity. }
\label{flares_nmus} 
\end{figure*}

\begin{figure*}
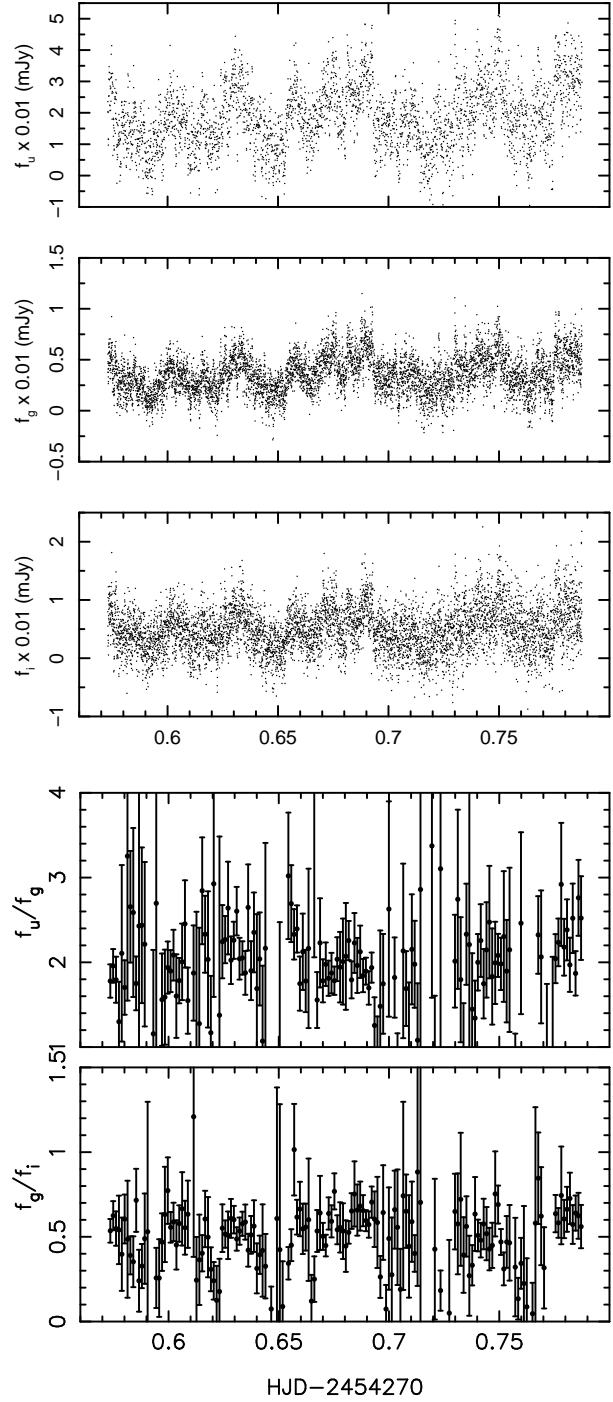

  \psfig{angle=0,width=8cm,file=shahbaz_fig3a.eps}

  \vspace{5mm}
  \psfig{angle=0,width=8cm,file=shahbaz_fig3b.eps}
  \caption{The top three panels show the \cen\ flare de-reddened flux density
\sloanu\, \sloang\ and \sloani-band lightcurves obtained by
subtracting a fit to the lower-envelope of the lightcurves shown in
Figure \ref{lcurve}. The  uncertainties in the \sloanu, \sloang\
and \sloani\ lightcurves are   1.2$\times 10^{-3}$\,mJy, 2.7$\times
10^{-3}$\,mJy and  7.0$\times 10^{-3}$\,mJy respectively. The bottom
two panels  show the flux ratio lightcurves binned to a time
resolution of 135\,s for clarity. }
  \label{flares_cen} 
\end{figure*}

\begin{figure*}
  \psfig{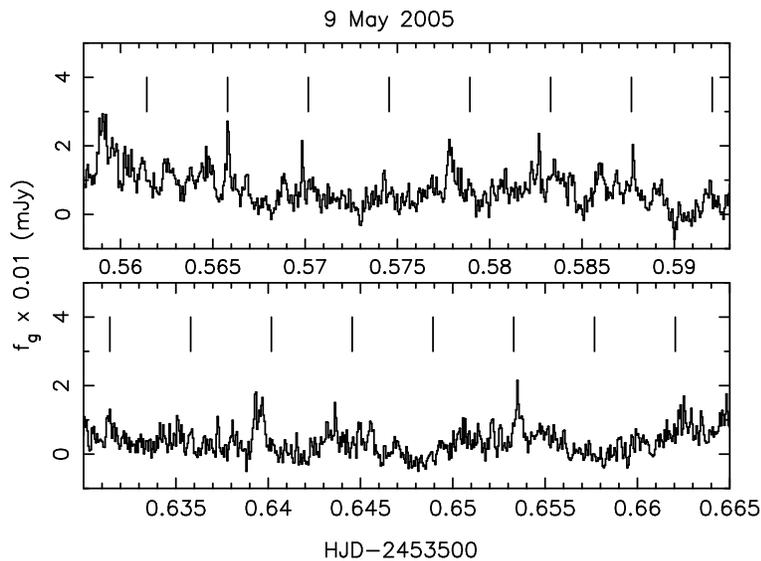}
  \caption{A close-up of the \sloang-band lightcurve of \nmus. 
  Short-term flare events seem to
  be periodic, but only over a few cycles.
  Vertical tick marks are shown with a representative interval of 6.3\,min. }
  \label{pulse}
\end{figure*}

\begin{figure*}
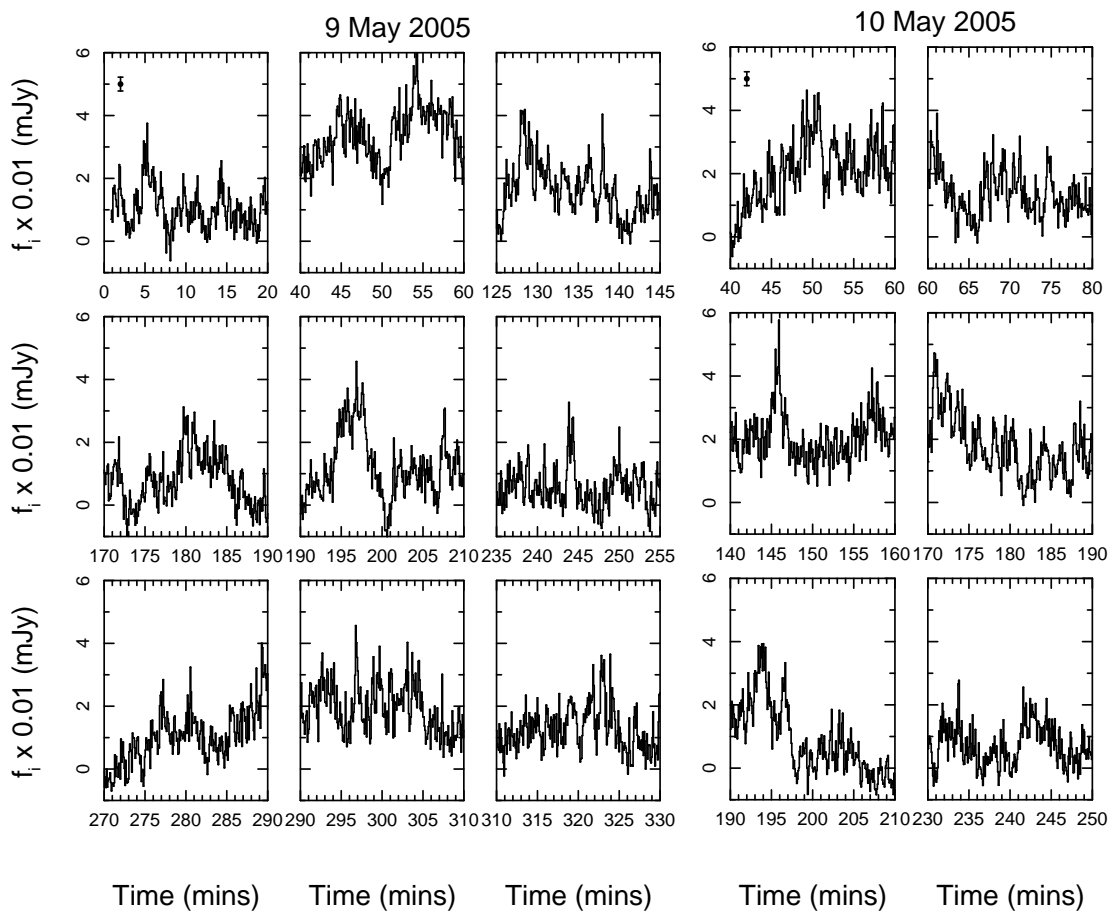

  \psfig{angle=0,height=12.0cm,file=shahbaz_fig5a.eps}
  \psfig{angle=0,height=12.0cm,file=shahbaz_fig5b.eps}
  \caption{Detailed plot of some of the individual flares in the
\sloani-band lightcurve of \nmus. Note that numerous flare events which
last for a few minutes are not resolved. 
The solid point in the top left panels 
marks the typical uncertainty  in the data. }
  \label{profiles}
\end{figure*}

\clearpage

\clearpage

\clearpage

%

\end{document}